\documentclass[numreferences]{kluwer}
\usepackage{amsmath}
\usepackage{mathptmx}
\usepackage{times}

\def\b{\bigskip}
\def\no{\noindent}

\def\Rit{\mathbb{R}}
\def\Cit{\mathbb{C}}
\begin{document}
\begin{article}
\begin{opening}
\title{Retrospective on Quantization}
\author{Christian Fr\o nsdal} 
\institute{Department of Physics and Astronomy, University of California, \\
Los Angeles, CA 90024-1547, USA \email{fronsdal@physics.ucla.edu}}
\date{Received: December 5, 2000}
{81Txx (81-XX, 16E40)}
\runningtitle{\sc Retrospective on Quantization}
\runningauthor{\sc C. Fr\o nsdal}
\begin{abstract}
Quantization is still a central problem of modern physics. 
One example of an unsolved problem is the quantization of Nambu mechanics.
After a brief comment on the role of Harrison cohomology, this review
concentrates on the central problem of quantization of QCD and, more
generally, quark confinement seen as a problem of quantization. Several
suggestions are made, some of them rather extravagant.
\end{abstract}
\end{opening}
\section{Quantization}

A well known text book on Quantum Mechanics offers, in an early chapter, 
a  Universal Quantization Paradigm. After the usual familiar applications 
one finds, towards the end of the book, a chapter dealing with the rigid
rotator. Needless to say, the ``universal" paradigm has been forgotten.

I remember trying to explain to a colleague, more than 25 years ago, 
what we were trying to understand about quantization. He would not agree 
that there was a problem.
Of course, we were all aware of the connection between
quantization and the theory of group representations; this was
explained, in principle, already by von Neumann, in
his proof of the uniqueness of the unitary representation of
the Heisenberg group. But few people anticipated the greatly
expanded role that has come to be played by the Heisenberg group
in modern mathematics, or the much more sophisticated applications
of group representation theory found in modern physics.

Mosh\'e had come to realize that the essence of quantization is
deformation theory, almost any deformation of a classical
structure amounts to quantization  \cite{Fl1}. Perhaps the most radical
example turned up during his last attack on the problem of
quantization of Nambu mechanics  \cite{DFST}. This problem leads to a
search for an \textsl{abelian} deformation of the ordinary
algebra of functions. It turns out that such deformations are
governed by Harrison cohomology, just as non-abelian deformations
are  described within de Rham cohomology. But Harrison cohomology
is trivial on smooth manifolds \cite{Pi}. This is an interesting difficulty,
but a greater obstacle is the fact that few mathematicians have
taken up the study of this subject, so that no instructive
examples of Harrison cohomology are available \cite{H, Ba, GS}.   
The following example may therefore be of some interest.

Consider the algebraic variety $A$ (a cone, or
degenerate conic) associated with conformal field theory in
$n-1$  real, Euclidean dimensions,
$$
A = \Rit^{n+1}/\sum_{i = 1,...,n}x_i^2-y^2=0\,.
$$
Let $W$ be the commutative $\Cit$-algebra of polynomials in the
coordinates $x_1,...,x_n,y$,
$$
W = \Cit[x_1,...,x_n,y],
$$
and let $W(A)$ be the quotient algebra
$$
W(A) = W/ \sum_{i = 1,...,n}x_i^2-y^2 = 0
$$
of functions on $A$. This algebra admits non trivial Harrison
cohomology. We exhibit a particular, non trivial cocycle and
use it to construct an abelian deformation of the algebra
of functions on $A$.

Every $f\in W(A)$ has a unique decomposition,
$$
f(x,y) = f_1(x) + yf_2(x),
$$
with $f_1,f_2$ being polynomials in $x_1,...,x_n$.
Define a deformed product, denoted $*$, by
$$
f*g = fg + \lambda C(f,g),~~ C(f,g) := f_2 \,g_2,~~ f,g \in W(A).
$$
That $C$ is a cocycle is readily verified; that is,
$$
dC(f,g,h) := fC(g,h) - C(fg,h) + C(f,gh) - C(f,g)h = 0.
$$
It follows that, to first order in $\lambda$ (regarded as a
formal variable) the deformed product is associative.
(Actually, it is associative to all orders in $\lambda$.)
That it is not trivial can be seen by the observation that,
after fixing
$\lambda$ in
$\Rit$, it is isomorphic to
$$
\Cit [x_1,...,x_2,y]/\sum_{i = 1,...,n}x_i^2 -y^2 + \lambda
= 0,
$$
an algebra of functions on the nondegenerate conic
$\Rit^{n+1}/\sum_{i = 1,...,n}x_i^2 -y^2 + \lambda = 0$.

I suggest that Harrison cohomology may have interesting
applications in quantum field theory, especially in connection
with certain anomalies, and perhaps in the context of operator
product expansions.

\section{Color}
The success of the concept of color in strong interaction
physics goes far beyond solving the problem for which it was
invented, that of allowing quarks to evade the constraint that
connects spin and statistics \cite{T,G-M}. It is a pity, nevertheless,
that this rather primitive idea of color has displaced the far
more interesting concept of parastatistics.  Wigner's point of
view, that led him to invent parastatistics, is curiously close
to the more modern concept of current algebra \cite{Wig}. Evidently,
the observables are the currents. In conventional local quantum
field theory they are bilinears in the enveloping algebra of a
Heisenberg algebra; their algebraic structure is that of
$Sp(\infty)$, and if the linears and the unit are included we
obtain the super Lie algebra
$OSp(\infty)$. Naturally, one is interested in
representations of this algebra, more particularly in those of
highest weight, realized in Fock spaces. The familiar ones
are induced from one-dimensional representations of the
subalgebra that is generated by $1, a_i$ and $a_ia^*_j +
a^*_ja_i,~i,j = 1,2,...~$, of the form
$$
a_i|0\rangle = 0~~(a_ia^*_j +a^*_ja_i)|0\rangle =
\lambda\,\delta_{ij}|0\rangle.
$$
The representation, and with it the statistics, is completely
fixed by the number $\lambda$. This is a simple but fundamental
fact that deserves to be mentioned in any text that aims to
explain Bose-Einstein quantization. The simplest case is
$\lambda = 1$; in this case one easily verifies that the vector
$$
(a^*_ia^*_j -a^*_ja^*_i)|0\rangle
$$
has zero norm. To get a Hilbert space, one must project out the
subspace that contains all null vectors, hence the postulate
that creation operators commute is actually redundant, as is the
statement that the many-particle states are symmetric. Higher,
integer values of $\lambda$ lead to parastatistics  \cite{GM,FF1}.

Interest in parastatistics had another aim in the
context of strong interactions, the problem of confinement
\cite{GM}. Taken in a very general sense, the problem is to
construct a theory that contains particles that do not
appear as asymptotic states. This is very easy to do in a
nonrelativistic context, but very difficult to reconcile with
relativity and the precepts of local field theory. Let us
talk  about this extremely important subject.

The hypothetical theory that I want to discuss contains
field operators that have some properties in common with the
quark fields of QCD. Namely, they have the same formal
transformation properties under various groups, and they
appear in perturbation theory in much the same way. But the
crucial properties are these:

\medskip
(1) The theory contains no
asymptotic quarks and

\smallskip
(2) the asymptotic states
are (in some sense) systems consisting of several quarks.

\medskip
\no Here
are some of the ways that one can fantasize about these objects.

A. The quarks are much like ordinary particles, but they are
prevented from separating from each other because they are
tied together with rubber bands or flux tubes. This is the
most ``physical" model. The most basic properties of the theory
are very different from the theory of free quarks, and the true
nature of the theory is not revealed by perturbation theory.

B. The quarks are singletons; quantum fields that are not
locally observable except when produced in pairs \cite{FF2}.
Ordinary massless fields are singleton currents, quantization
relies on a generalized version of the Heisenberg algebra. The
quark fields are permitted to be non-local since they are not
locally observable. Indeed, it seems a pity, when one tries to
deal with unobservable fields, not to take advantage of this
freedom to step outside the narrow framework of local field
theory\footnote{Take note, however, that a resolution
of the ills of QCD will be worth a million dollars US only
if found strictly within the conventional framework,
unlikely as this is.}.   Singleton field theory
provides an example of  topological  gauge fields that can be
transformed away locally; possibly there are other ways to
introduce quark fields that are artifacts of a gauge
symmetry. The problem is to relate them to hadrons.

C. Perturbative local field theory is somewhat at a loss when it
comes to incorporating unstable particles. It is possible,
however,
to concoct a real Lagrangian that describes particles with complex
masses,  that I shall call quirks, interacting with more
conventional fields, and such that the Fourier transforms of the
propagators are analytic near the real axis. If $\psi(p)$ is
the Fourier transform of such a field operator, then the norm of
the asymptotic state associated with it is given by the
discontinuity of the generalized function $G(p^2)$ defined by
$$
 \langle 0|\psi(p)\bar\psi(q)|0\rangle = \delta(p-q)G(p^2)
$$
across the real axis, and this vanishes for quirks. Hence states
that contain an isolated quirk is null. States that contain two or
more quirks do not have zero norm; or rather, whether they
do or do not depends on the Feynman rules that one invents for
them. More about this below.

D. Recall that the scattering amplitudes of string theory,
invented by Veneziano  \cite{V}, describe the scattering of
spinless particles that turned out to be tachyons. String
theory came of age when, with the aid of supersymmmetry, the
means were found to decouple this unphysical mode from the
physical states. Yet all string states are constructed as
states of two or more of these objects. The similarity to
the philosophy of QCD is striking. The ``constituents" of
string theory are bosons, those of QCD  are fermions. What
is wanted is a QCD without physical quarks, or a string
theory with constituents that have some  of the properties
of quarks. This topic will also be developed below.

\section{Quirks}

 The dominant modern attitude to elementary particle physics is
based on the idea of scattering matrix, a mapping of a Hilbert
space of initial on an identical Hilbert space of final states. The
salient point about unstable particles is that they do not appear
among these states. Consider a fictitious unstable 
particle~--~a ``quirk"~--~ and let us try to associate a
fundamental field operator to it. If all interactions are switched
off, then the unstable particle presumably becomes stable, and in
that limit it must appear in the Hilbert space of asymptotic
states; at least, that is the conventional point of view. But this
would imply an important discontinuity in the structure of the
space of asymptotic states, and consequently a discontinuity of the
values of  physical observables, as functions of the coupling
strengths;  the  very concept of analytic perturbations becomes
meaningless.

The suggestion that we offer here is as follows. Let us incorporate
the unstable nature of quirks into the point of departure of
perturbation theory. In the free theory there are stable particles
represented by free fields, but the free Lagrangian also contains
fundamental noninteracting quirk fields. Once we understand
\textsl{this} free field theory, then interactions can be
introduced, and now there is a hope of constructing an analytic
perturbation theory. Procedures such as this are familiar in the
context of quantum mechanics, where the  first step in setting up
a perturbation scheme consists of splitting the Hamiltonian into two
parts. Two considerations govern the splitting; that is, the choice
of the unperturbed Hamiltonian and {\it ipso facto} the nature of
the unperturbed system: 
(1) The free theory should be exactly solvable, and 
(2) the perturbation should be analytic.
(Example: a Schr\"odinger atom with an additional potential
proportional to $1/r^2$; it is necessary to include this term in the free
Hamiltonian. Another example: Dollard's treatment of the infrared
problem \cite{Do}.) The use of counterterms in renormalization
theory can be described in the same way. And here the
analogy is a close one, for the width of the resonance may
well be renormalized in perturbation theory. The shift of
the vacuum in the Standard Model, to the minimum of the Higgs
potential, offers another close analogy,  for the
non-vanishing vacuum expectation value redefines the free
theory by incorporating into it a feature of the
interaction, to render the perturbation analytic (or at
least, more likely to be analytic).

Let $ |0\rangle$ be the unique vacuum state, and let $\psi$ denote
a free Dirac-type quirk field operator. We anticipate a
perturbation theory in which the calculation of S-matrix elements
reduces to the evaluation of vacuum expectation values of products
of field operators. Suppose that
$$
\psi(f)|0\rangle \neq 0,~~~\psi(f) = \int f(x)\psi(x),
$$   then there is no alternative to accepting the fact that this
state must have a place in the theory. But there is no need to jump
to the conclusion that it belong to the Hilbert space of asymptotic
states.  Taking a cue from gauge theories, we suppose that
fundamental field operators act in a larger space, possibly
indefinite, and that the physical Hilbert space is a quotient space
(or a quotient of a semi-definite subspace) by a subspace of null
states (states of vanishing norm). In particular, to avoid the
appearance of quirks in the asymptotic states, we must suppose that
the asymptotic state associated with $\psi(f)|0\rangle$ have
zero norm.

Let us further assume the conventional construction of the inner
product, in terms of $\psi(f)$ and a hermitian conjugate field
operator $\bar \psi(f)$, namely, the norm of this asymptotic
state is determined by a discontinuity of the matrix element
$\langle 0|~ \bar\psi(f)
 \psi(f)~|0\rangle.$
Unitarity of the theory demands that this discontinuity be represented as
$$
\sum_n \langle 0|~ \bar\psi(f)  |n\rangle\langle n|
  \psi(f)~|0\rangle,
$$
where the sum runs over a set of asymptotic states. Since the quirk
field operator does not create such states, the norm must be zero:
$$
{\mathrm{Discontinuity~}} \langle 0|~ \psi(f) \bar\psi(f)~|0\rangle = 0.
$$

This is our main premise: that a state created from
the vacuum by the quantum field operator associated with an
unstable particle  have zero norm. Let us emphasize that this
refers to the action of {\bf one} fundamental, unstable field
operator; we shall see that some states containing two unstable
particles do not have zero norm. This is a minor miracle that
reflects the fact that, strictly speaking, quantum field operators
are not operators (as we know, for example, by the existence of anomalies).

\section{The quirk Lagrangian}

We construct an example of a Lagrangian field theory of
noninteracting quirks. Since the Lagrangian must be real we
need two spinor field, $\psi_1$ and $\psi_2$. The following vacuum
expectation values for quirk fields incorporates all expected properties,
\begin{eqnarray*}
\langle 0|\psi_1(x)\bar \psi_2(x')|0\rangle &=& \int dp~ e^{ip(x-x')}
\frac{1}{p-m +i\lambda},\\
\langle 0|\psi_2(x)\bar \psi_1(x')|0\rangle &=& \int dp~
e^{ip(x-x')}  \frac{1}{p-m -i\lambda},
\end{eqnarray*}
where $m$ and $\lambda$ are real parameters that characterize the
quirk. (The integration runs over $\mathbb{R}^4$.)
Namely, (1) it has the usual hermiticity property for Dirac
fields,~ (2) the Fourier transform can be continued to a function
analytic near the real axis, with no discontinuity, and
consequently  the norm of the asymptotic state associated with
$\psi(x)|0\rangle$ is zero, (3) the matrix elements fall off
exponentially for large values of $|(x-x')^2|$.

This function will take the place, for quirk fields, of the usual
vacuum expectation value of the time ordered product. To complete
the Wick rules we need to define the vacuum expectation values of
products of the fundamental fields; we may assume that this will
involve symmetrization, and it is immediately clear that there
may be an opportunity for getting around the usual connection
between spin and statistics.

A real Lagrangian that is consistent with these rules is,
$$
L = \bar\psi_1(p-m+i\lambda)\psi_2  - \bar\psi_1J_1 -
\bar\psi_2J_2 + {\rm h.c.}~.
$$
A change of basis leads to
$$
L = \bar\psi_+(p-m)\psi_+ - \bar\psi_-(p -m)\psi_- +
\lambda(\bar\psi_-\psi_+ + \bar\psi_+\psi_-)  -(\psi_+J_+ +
\psi_-J_- + h.c.).
$$
The $\lambda$-term can be introduced via a Yukawa interaction 
with a scalar field that is given a non-vanishing vacuum expectation value.  
Of course, the $m$-term can be interpreted in the same way, by the usual 
Higgs mechanism. It is perhaps significant that both sectors are unstable  
without the vacuum expectation value.

The idea of associating certain exotic particles with null states
has been advanced in other contexts as well, as we have mentioned
already.   The most obvious apparent difficulty would seem to be
the argument that ``if $\psi|0\rangle$ has zero norm, then surely 
$\psi\psi|0\rangle$ must have zero norm as well".  This would be true 
if the operator products were not singular. The norm of a 2-quirk state; 
for example, of the asymptotic state associated with
$\psi_1(f)\psi_2(f')|0\rangle$, is determined by the discontinuity of
$$
\int dk  \frac{1}{k  -m +i\lambda } ~\frac{1}{p-k  -m -i\lambda }
f(k)f'(p-k).\eqno(4.1)
$$
Ordinary Feynman rules, applicable to stable particles, are
evaluated by conversion to complex contour integrals, and
deformation of the contours of integration over internal energies,
here $k_0$. The location of the singularities, in the second and
fourth quadrants, authorizes the rotation of the contour from the
real axis to the imaginary axis. (More precisely, this is true for
real $\vec k$ and for $p$ real and spacelike.) In
(4.1) we have an integrand with poles in all four quadrants of the
complex $k_0$ plane, and rotation of the contour is not
possible.   When $\lambda$ is not zero, and if the $k_0$ contour
follows the real axix, then this function has no discontinuity
for any real value of $p$. Instead we shall define the two-point
quirk correlation function by (4.1), but with the $k_0$
integration following the imaginary axis.
  The usual treatment of this integral leads to
an expression that is analytic in $p$ except for singularities of a
parameter-integral of the form
$$
\int_0^1d\alpha \big[ \alpha^2p^2 - \alpha(p^2 - 4i\lambda m) +
(m-i\lambda)^2\big]^{-2}.
$$
Vary $p^2$ along the real axis, from left to right. The two poles
start at
$\alpha = 0,1$, move into the upper half plane until they meet at
the point $(1 + im/\lambda)/2$ when $p^2 = -4\lambda^2$. One of
them goes to infinity as $p^2$ passes the origin and returns in the
lower half plane, while the other crosses the real axis at $1/2$
when $p^2 = 2m^2-2\lambda^2$. The contour of integration  finally
gets pinched at at $(1 -i\lambda/m)/2$ when $p^2$ reaches the value
$4m^2$. Result: there is branch point at $p^2 = (2m)^2$,
independently of the value of $\lambda$.

Thus we learn that the asymptotic state   associated
with $\bar\psi_1\bar\psi_2 |0\rangle$ does not have zero norm, in
spite of the fact that the states
$\bar\psi_1 |0\rangle$ and $\bar\psi_2|0\rangle$ are null states.
There are physical, propagating states with masses above $2m$,
independently of the value of $\lambda$.

\section{Dynamics of string theory}

Now I want to tell you a
very ancient story, as briefly as I can. The story begins with the
invention of the Dirac equation  \cite{Di},
$$
(p\cdot \gamma - m)\psi(x) = 0.
$$
Recall that $\psi$ is a field on space time, taking values in a
finite dimensional Lorentz module. Solutions exist only for
timelike momenta, this is of primary importance to the
interpretation. Dirac's equation applies to electrons; for a
while it was thought to apply to protons as well. Of course it
does, in a very approximate sense. But to keep to the historical
order let me next recall the very interesting
work of  Majorana  \cite{M}.

 Majorana may have had the same motivation as Dirac,
but he was less ready to accept the presence of solutions with
negative energy. He asked if it is possible for the operator
$\gamma_0$ to have a strictly positive spectrum. Of course,
relativistic invariance demands that the field take values in a
Lorentz module, and the answer found by Majorana was that yes, it
is possible if this module is unitary and hence
infinite dimensional. Majorana's work was ignored.

I come now to the applicability of Dirac's equation to the proton.
As the experimental probing of proton structure advanced, it
became clear that it was not a simple Dirac particle. The
structure was described in terms of elastic form factors. These
form factors decrease with increasing momentum transfer, and this
is not consonant with Dirac's equation. More generally, any
particle described by a Dirac-type equation, where the field lives
in a finite dimensional Lorentz module, has polynomial form
factors. In contrast, the form factor of a Majorana particle does
decrease with increasing momentum transfer; in fact, it looks
very much like the proton form factor found experimentally.

For this reason, theories of the Majorana type became,  briefly,
quite popular. One knows what is wrong with these
infinite component field theories, but let me first go over some
of their good points. Perhaps  the most surprising fact is that
some theories of this type have a very sensible physical
interpretation, as two-particle systems  \cite{Fr1}. In fact, one
of the useful applications is a marked improvement of the
Bethe-Salpeter method for dealing with bound states in
quantum field theory~\cite{FH}. The non-relativistic limit is
 instructive, it is a formulation of the theory of the
Schr\"odinger hydrogen atom. Of course, here we shall not see
a Lorentz module, but a Galilei module. What is important is
that we are describing a system with a complicated internal
structure. I shall limit myself to discussing just one
aspect of the theory, the structure of perturbation theory.

Consider the scattering of photons from an atom  \cite{Fr2}. The
Feynman diagram represents a sequence of events taking place:
 incoming photon and atom, absorption of the photon, propagation
of the atom in an excited state, emission of a photon,
outgoing photon and atom. The amplitude for the process is a
parallell sequence: a vertex factor $\exp(i\vec k\cdot \vec r)$, a
propagator $(E-H_0)^{-1}$, a vertex factor 
$\exp  (-i\vec k'\cdot \vec r)$. The propagator depends on the energy 
$E$ of the intermediate state, but it is not sensitive to the momentum.
This is because the momentum of a very heavy object is not
observable. But we may, if we wish, define the hamiltonian of
an atom with momentum $\vec k$ by setting
$$
H_{\vec k}:= \exp(i\vec k\cdot \vec r)H_0\exp(-i\vec k\cdot \vec
r).
$$
The Schr\"odinger equation now involves $\vec k$ and so does the
wave function. In this  entirely equivalent formulation of the
problem ... and this is the nonrelativistic approximation to
Majorana's theory ... the  amplitude is constructed in terms that
are those of relativistic local field theory, namely: Incoming
photon with momentum $\vec k$ and atom with momentum $\vec p$, (no
exponential factor), propagator for an atom with momentum
$\vec k+\vec p$, (no exponential factor), outgoing photon with
momentum $\vec k'$ and atom with momentun $\vec p+\vec k-\vec k'$.

String theory took over the main ideas of infinite component
field theories. The spark of the string theory revolution came
from the realization that  pole dominated scattering amplitudes
behave better when there is an infinite sequence of poles. But
the known infinite component field theories suffered from the
ubiquitous space like solutions, about which more below; that is,
bad behaviour with respect to crossing symmetry. The additional
ingredient of string theory is duality: one contrives to
construct an amplitude that is at the same time a sum over pole
contributions in the direct channel and a sum over pole terms
in the crossed channel. This was accomplished by Veneziano~\cite{V}.

In the beginning there was just Veneziano's amplitude, and the 
multiparticle generalization. Then it was shown that these amplitudes 
could be factorized, just like the amplitude that we have been talking 
about, into a sequence of propagators separated by 
exponential momentum factors  \cite{FGV}.
But what followed created a huge gulf between string theory
and Majorana theory. Fubini and Veneziano  \cite{FV}, in a very
beautiful paper, succeeded in doing exactly the opposite of
what I just did for the hydrogen atom. Instead of getting
rid if the vertex factors, they managed to elimate all the
propagators, thereby creating vertex algebras and two
dimensional quantum field theory!

This result is truly wonderful, but   something was  lost.
The well known Lagrangian string theories in
high dimensions describe the dynamics of  elementary boson and
fermion fields. But the dynamical fields of string theory are the
vertex operators, exponentials of these elementary fields;
scattering amplitudes are constructed entirely out of vertex
operators. There may be something like a string field operator,
directly related to vertex operators, and we do have some hints
about the dynamics, as we shall see.

So far I have stressed only the attractive features of Majorana's infinite
dimensional version of Dirac's equation. The bad news is that it
possesses solutions with space like momenta. Similar equations,
studied intensively during the late sixties, including
relativistic versions of the hydrogen atom, all shared this
devastating feature. There was a time when theories were said to
be implausible to the degree that they predict the existence of
new particles. The Dirac equation predicted only the positron,
and that was not too much, but the Majorana equation and its
generalization predicted infinite numbers of excited particle
states. Actually, this feature should be counted an advantage,
especially after the advent of dual models and the realization
that it was responsible for the good properties of form factors.
The irony is that the predicted excited states turned out
to be too few!

Most models have energy spectra labelled by an integer, $n$ say.
The degeneracy of the $n$'th level, in theories of the Majorana
type, are polynomial in $n$. In dual models the degeneracy grows
exponentially. This may be an essential characteristic of theories
that avoid spacelike solutions. The degree of degeneracy
associated with the Veneziano amplitude is known from the work of
Fubini and Veneziano \cite{FV} on factorization of this amplitude. In
this work there appears a propagator, namely
$$
(L_0- p^2)^{-1},~~ L_0 = -2(\sum na^*_na_n-1),
$$
where $L_0$ is the operator that was later interpreted as the
most important operator of the Virasoro algebra, $\{a_n\}$
is an infinite set of oscillator operators and $p$ is the
momentum of the state. Let $|0\rangle$ be the vacuum associated
with all these oscillators, then the domain of $L_0$ is the space
$[[a*_1,a^*_2,... ]] \otimes |0\rangle$, and the mass$^2$
spectrum of the model is the spectrum of $L_0$. In other
words, the states satisfy the equation
$$
( L_0-p^2)\Psi = 0,~~ \Psi \in [[a_1^*,a^*_2,...]] \otimes
|0\rangle.
$$
Though second order, this equation is very much of the Majorana
type, though the internal space is bigger than anything
that Majorana or his emulators had dreamed of.

This equation is a big part of a dynamical
theory of strings. It certainly does not contain everything, for
there is one (only one!) spacelike solution, and plenty of states
with negative norm. Some of the  states of negative norm are
eliminated from the Veneziano amplitude by the exponential factors
associated with the vertices. Fubini and Veneziano showed that
the amplitude could be understood entirely in terms of vertex
operators, after absorbing the propagators into the vertex
operators. My inclination is to try the opposite tack, to improve
the propagator by incorporating into it the role of the
vertex factors. This should facilitate the task of constructing
the interaction.

A promising attempt to formulate an action principle for
quantum string dynamics was made by Witten 15 years ago \cite{Wit}.
This work seems to be applicable only to open strings. I
believe that this work should be taken up again, but not with the
too
modest aim of incorporating closed strings. If the result is
to have any bearing on the outstanding open problem of our time,
then it is not enough just to look for an extension of
Witten's treatment of the Veneziano model; what is needed is
a very nontrivial generalization.

\section{Summary}

Quantization still offers challenging mathematical problems,
but the biggest problem of all is to make sense of QCD. There
are hints that string theory, more precisely the Veneziano
model, may point in the right direction. The suppression of the
tachyon in super string theory suggests that quarks (and gluons)
may be the suppressed constituents of a more elaborate version
of the Veneziano model. Steps towards such a generalization
could include the following small steps:

(1) Incorporate the Fadde'ev-Popov ghosts into the vertex
operators. (2) Construct the vertex operators for the
superstring. (3) Ditto on an AdS background.  (4) Replace the outer,
scalar fields in the amplitude by scalar superfields. (Note that
there are no supersymmetric tachyons!) (5) Combine (3) and (4).
 Possibly, the first
two may have been accomplished already  \cite{BV,DW}.
The third, without ghosts and without supersymmetry, is easy. For
the 4-point function the result is
$A(s,t)\Psi(x_1,x_2,x_3,x_4)$, evaluated at $x_1 = x_2, x_3 =
x_4$, where $A$ is Veneziano's function, $s = (\bigtriangledown_1
+ \bigtriangledown_2)^2,~ t = (\bigtriangledown_2 +
\bigtriangledown_4)^2$ are covariant d'Alembertians and $\Psi$ is
a product of tachyon wave functions. The fourth suggestion
seems to lead to a spectrum $m \propto n$ (instead of $m^2
\propto n$). Finally the fifth one may include an interesting
possibility involving singletons and massless particles with
all spins.

\b %\b
\no{\bf Acknowledgements.}

I thank Murray Gerstenhaber for introducing me to Harrison
cohomology, Georges Pinczon for the suggestion to study
functions on cones, and Dirk Kreimer, Daniel Sternheimer and
Ivan Todorov for useful criticism.
%\newpage

\end{article}

\begin{thebibliography}

\bibitem{Ba} 
Barr, M., ``A cohomology theory for commutative algebras. I, II",
Proc. Amer.  Math. Soc. {\bf 16} (1965), 1379--1384. \\% Barr, M.
``Harrison Cohomology, Hochschild Cohomology, and
Triples", J. Algebra {\bf 8} (1968) 314--323.

\bibitem{BV} 
Berkovits, Nathan; Vafa, Cumrun. ``$N=4$ topological strings". 
Nuclear Phys. {\bf B 433} (1995) 123--180 (\texttt{hep-th/9407190}).

\bibitem{Di} 
Dirac, P.A.M., Proc. Roy. Soc. {\bf A 167} (1938), 148.

\bibitem{DW} 
Dolan, Louise; Witten, Edward. ``Vertex operators for AdS$_3$ background
with Ramond-Ramond flux". J. High Energy Phys. 1999, no. 11, Paper 3, 21pp.
(\texttt{hep-th 9910205}).

\bibitem{DFST} 
Dito, G.; Flato, M.; Sternheimer, D. and Takhtajan, L.
``Deformation quantization and Nambu mechanics",
Comm. Math. Phys. {\bf 183} (1997), 1--22 (\texttt{hep-th/9602016}).

\bibitem{Do} 
Dollard, John D. ``On the definition of scattering subspaces in 
nonrelativistic quantum mechanics. J. Mathematical Phys. {\bf 18} 
(1977), 229--232. \\%Dollard, John D. 
``Asymptotic convergence and the Coulomb interaction".
J. Mathematical Phys. {\bf 5} (1964) 729--738.

\bibitem{Fa} 
Faddeev, L.D.,  ``Operator anomaly for the Gauss law",
Phys.Lett. {\bf B 145} (1984) 81--84.

\bibitem{Fl1} 
Flato, M.,``Deformation view of physical theories'',
Czech.J.Phys. {\bf B 32}  (1982), 472--475.

\bibitem{FF1}  
Flato, M. and  Fronsdal, C., ``Parastatistics, highest weight 
$\mathfrak{osp}(N,\infty)$ modules, singleton statistics and confinement",
J.Phys.Geom. (1989), 293--309.

\bibitem{FF2} 
Flato, M. and  Fronsdal, C., ``Quarks or Singletons",
Phys.Lett. {\bf B 172} (1986) 412--416.

\bibitem{Fr1} 
Fronsdal, C., ``Infinite multiplets and the hydrogen atom",
Phys. Rev. {\bf 156} (1967), 1665--1677.

\bibitem{Fr2} 
Fronsdal, C., ``Compton scattering from bound electrons",
Phys. Rev. {\bf 179} (1969), 1513--1517.

\bibitem{FH}   
Fronsdal, C. and Huff, R.W., ``Two-Body Problem in Quantum
Field Theory", Phys.Rev. {\bf D3} (1971), 1689.

\bibitem{FGV}  
Fubini, S.,  Gordon,  D.and Veneziano, G., ``A General Treatment 
of Factorization in Dual Resonance Models", 
Phys.Lett. {\bf 29B} (1969) 679--82.


\bibitem{FV}  
Fubini, S. and Veneziano, G., ``Duality in operator formalism", 
Nuovo Cim. {\bf 67A} (1970), 29-47.

\bibitem{G-M} 
Gell-Mann, M., ``Current topics in particle physics", in
{\it Proceedings of the thirteenth International Conference on
High-Energy Physics}, Berkeley 1966.

\bibitem{GS}    
Gerstenhaber, M. and Schack, S.~D., ``A Hodge type
decomposition for commutative algebra cohomology", J. Pure and
Appl. Alg. {\bf 48} (1987) 229-247. 
%Gerstenhaber, M. and  Schack, S.~D.,
``Algebraic cohomology and deformation theory.
Deformation theory of algebras and structures and applications," 
(Il Ciocco, 1986), 11--264, NATO Adv. Sci. Inst. Ser. C: Math. Phys. Sci., 247,
Kluwer Acad. Publ., Dordrecht, 1988.

\bibitem{GM} 
Greenberg, O.~W., ``Spin and unitary-spin independence in a
paraquark model of baryons and mesons", Phys.Rev.Lett.  {\bf 13}
(1964), 1100-1102. %and Messiah, A.M.L., ``Selection rules for
%parafields and the absence of of para-particles in nature",
%J.Math.Phys. {\bf 6} (1965), 1155-1167.

\bibitem{H}  
Harrison, D.~K., ``Commutative Algebras and Cohomology",
Trans. Amer. Math. Soc. {\bf 104} (1962), 191--204.

\bibitem{M} 
Majorana, H., Nuovo Cimento {\bf 9} (1932), 335.

\bibitem{OK} 
Ohnuki, Y, and Kamefuchi, S., ``Wave functions of identical
particles",  Ann.Phys.  {\bf 57} (1969) 337--358.

\bibitem{Pi} 
Pinczon, Georges. ``On the equivalence between continuous and differential 
deformation theories. Lett. Math. Phys. {\bf 39} (1997), 143--156. 

\bibitem{T}
Tavkhelidze, N., ``Higher Symmetries and Composite Models of Elementary
Particles", Proceedings of the Conference on High-Energy Physics and
Elementary Particles, International Centre for Theoreticle Physics, Trieste
1965.

\bibitem{V} 
Veneziano, G., ``Construction of a Crossing-symmetric,
Regge-Behaved Amplitude for Linearly Rising Trajectories", Nuovo
Cimento {\bf 57A} (1968) 190.

\bibitem{Wig} 
Wigner, E.~P.,``Do the equations of motion determine the
quantum mechanical commutation relations?", Phys. Rev. {\bf 77},
(1950), 711--712.

\bibitem{Wit} Witten, Edward. ``Noncommutative geometry and string theory",
Nuclear Phys. {\bf B268} (1986), 253--294.

\end{thebibliography}
\end{document}